\documentclass[aps,prl,amsmath,amssymb,preprint,groupedaddress,superscriptaddress]{revtex4-1}
\usepackage{graphicx}
\usepackage{amsmath}
\usepackage{amssymb}
\usepackage{mathrsfs}
\usepackage{epsfig}
\usepackage{color}
\def\comment#1{}

\begin{document}


\title{PSR J0337+1715: an appropriate laboratory for testing the Nordtvedt effect}


\author{Wen-Biao Han}

\email{wbhan@shao.ac.cn}

\affiliation{Shanghai Astronomical Observatory, Chinese Academy of Sciences \\
80 Nandan Road, Shanghai, China P.R., 200030}
\author{Shi-long Liao}
\email{shilongliao@shao.ac.cn}
\affiliation{Shanghai Astronomical Observatory, Chinese Academy of Sciences \\
80 Nandan Road, Shanghai, China P.R., 200030}
\affiliation{University of Chinese Academy of Sciences, Beijing , China P.R., 100049}

\date{\today}

\begin{abstract}
Highly accurate binary-pulsar timing plays an important role in test of General Relativity. In this Letter, we argue that PSR J0337+1715, a milli-second pulsar in a stellar triple system, could be a very good laboratory for testing the strong equivalence principle (SEP). From the reported orbital parameters of this triple system, we give an uplimited estimation of the  Nordtvedt parameter $\eta_\text{N} < 10^{-4}$ based on the orbital polarization calculation. This result is slightly better than the existed ones. In addition, if based on the observed uncertain value of inner orbital eccentricity, we even can obtain $\eta_\text{N}$ up to $10^{-7}$, which needs to be confirmed in future. However, more rigorous and accurate measurement of the Nordtvedt effect should be taken from more timing data of PSR J0337+1715.
\end{abstract}


\maketitle

\emph{Introduction} General Relativity --- the Einstein's theory of gravity, is now a base-stone of the modern gravitational theory, but still not proved completely. However, there are many theories try to challenge the status of General Relativity, for example, $f(R)$, MOND, scalar -tensor theory, and etc. Many experiments in the solar system (weak field and low velocity), and observations of strong field astrophysical phenomenon (binary pulsars, supermassive black holes and etc.) were made to test the gravitational theory \cite{Will06, Stairs03}.

One of the fundaments of General Relativity: the strong equivalence principle (SEP) can be checked by Lunar Laser Ranging (LLR) as proposed by Nordtvedt \cite{Nord68}. Then some people showed that using a group of pulsar-white-dwarf systems can also check the Nordtvedt effect based on probabilistic considerations \cite{Damour91, Wex97}. LLR tests have set a limit of $|\eta_\text{N}| \sim 10^{-4}$ \cite{Dickey94}, but pulsar-white-dwarf systems now only can give $10^{-2}$ uplimit, because of lack of such kind systems. In General Relativity, this value should be zero exactly.

More recently, a milli-second pulsar named PSR J0337+1715 was found in a hierarchical triple system \cite{pulsar3}. In detail, a binary composed by this pulsar with 1.4378 $M_\odot$  and a white dwarf with 0.19751 $M_\odot$ is orbiting around another white dwarf with 0.4101 $M_\odot$. This three-body system looks like our Earth-Moon-Sun system, but may more clean than the later because of without the perturbation from other planets. In addition, the gravitational interaction of the pulsar system is also much stronger than the Earth-Moon-Sun one.

In this Letter, for the first time, we show that this brand-new triple system, can be an appropriate laboratory to test the SEP. Using the published data in \cite{pulsar3}, we estimate a rough uplimit of the Nordtvedt parameter. However, more timing data in future can be used to constrain or measure this parameter directly. In the remain part of this Letter, we introduce the Nordtvedt effect briefly, then we calculate the orbital polarization of the compact triple system with a violence of SEP and give our estimation of $\eta_\text{N}$. Finally, conclusions and discussions are made in the end.

\emph{Nordtvedt effect}
The possibility of direct tests of the SEP through Lunar Laser Ranging (LLR) experiments was first pointed out by Nordtvedt \cite{Nord68}. As the masses of Earth and the Moon contain different fractional contributions from self-gravitation, a violation of the SEP would cause them to fall differently in the Sun's gravitational field. This would result in a  polarization of the Earth-Moon orbit in the direction of the Sun. For such kind three-body system, the equation of motion of relative position for the inner binary (e.g. Earth-Moon) is
\begin{align}
\boldsymbol{\ddot{r}}=-GM\frac{\boldsymbol{r}}{r^3}+\boldsymbol{a}_\text{R}+\Delta \boldsymbol{g},
\end{align}
where $M=m_1+m_2$. The first term in the right hand is the Newtonian gravitational force, the second one is the relativistic correction and the last one is due to the violation of SEP. In the third term, $\boldsymbol{g}$ comes from the third body 's gravitational force (e.g. the Sun), and $\Delta=\Delta_1-\Delta_2$. The ratio of gravitational and inertial mass is
\begin{align}
(\frac{m_\text{grav}}{m_\text{inert}})_i=1+\Delta_i \approx 1-\eta_\text{N}(\frac{E_g}{m c^2})_i,
\end{align}
where $i=1,2$, and $E_g$ is the negative of the gravitational self-energy of the body, $\eta_\text{N}$ is the so-called Nordtvedt parameter, and can be a combination of PPN parameters,
\begin{align}
\eta=\eta_\text{N}=4\beta-\gamma-3-\frac{10}{3}\xi-\alpha_1+\frac{2}{3}\alpha_2-\frac{2}{3}\varsigma_1-\frac{1}{3}\varsigma_2.
\end{align}

Then, from \cite{Damour91, Will1985}, as the Moon's self-graviational energy is smaller than the Earth's, the Nordtvedt effect causes the Earth and Moon to fall toward the Sun with slightly different accelerations:
\begin{align}
\delta a=\Delta g=\eta\left[\left(\frac{E_g}{mc^2}\right)_1-\left(\frac{E_g}{mc^2}\right)_2\right]GM_3/R^2
\end{align}
where $m_1$, $m_2$ and $M_3$ refers to the mass of the Earth, Moon and Sun. And $R$ is the distance between Earth and Sun.
The magnitude of $E_g/m c^2$ is $ 4.6 \times 10^{-10}$ for the Earth and $0.2 \times 10^{-10}$ for the Moon. And for the Sun is about $ 3.6\times 10^{-6}$; for the Jupiter $\sim 10^{-8}$; for neutron star $ \sim ~0.2$ and white dwarf  $ \sim 10^{-4}$.
%


Linearizing Eq.(1) about a circular orbit $(r=r_0+\delta r_\Delta)$, we obtain a polarization of the Earth-Moon system by the external field of the Sun \cite{Will1985}:
\begin{align}
\delta r_{\Delta}=\left(\frac{1+2\omega_\text{I}/ \Lambda}{\omega_\text{I}^2-\Lambda^2}\right)\delta a \cos{\Lambda t},
\end{align}
Where $\omega_\text{I}$ is the angular velocity of the Moon circle around the Earth, and $\Lambda=\omega_\text{I}-\omega_\text{O}$ with $\omega_\text{O}$ is the angular velocity of the Earth circle around the Sun.

\emph{Compact triple system} For the triple system PSR J0337+1715 we studied \cite{pulsar3}, it is a millisecond pulsar in a hierarchical triple system with two white dwarf companions. The gravitational field of the outer white dwarf strongly accelerates the inner binary containing the neutron star. This system is similar with the Moon-Earth-Sun system but with more strong gravitational interaction and much larger special gravitational self-energy (about 0.2 for neutron star, but for the Earth this value is only $10^{-10}$). Therefore PSR J0337+1715 will  provide an ideal laboratory for testing the strong equivalence principle of General Relativity. Some system parameters for PSR J0337+1715 are listed in Table I.

\begin{table}
\begin{center}
\begin{minipage}[]{110mm}
\caption[]{System parameters for PSR J0337+1715
 \label{tab1}}\end{minipage}
\setlength{\tabcolsep}{1pt}
\small
 \begin{tabular}{ccccccccccccccc}
  \hline\noalign{\smallskip}
Parameter& & && &Symbol&& &&&& Value \\
  \hline\noalign{\smallskip}
Pulsar period& &&&&$P$&&&&&&2.73258863244(9) ms\\
Pulsar semimajor axis (inner)& &&&&$a_\text{I}$&&&&&&1.9242(4) ls\\
Eccentricity parameter 1 (inner) & &&&&$(e\sin\omega)_\text{I}$ &&&&&& $6.8567(2) 10^{-4}$ \\
Eccentricity parameter 2 (inner) & &&&& $(e\cos\omega)_\text{I}$ &&&&&& $9.171(2) 10^{-5} $ \\
Pulsar semimajor axis (outer)& &&&&$a_\text{O}$&&&&&&118.04(3) ls\\
Eccentricity (outer)& &&&&$e_\text{O}$&&&&&&$3.53561955(17)\times10^{-2}$\\
Orbital period of Inner orbit& &&&&$P_\text{I}$&&&&&&$1.629401788(5)  $d\\
Orbital period of Outer orbit& &&&&$P_\text{I}$&&&&&&$327.257541(7) $d\\
Inclination of invariant plane& &&&&$i$&&&&&& $39.243(11) ^{\circ}$ \\
Inclination of inner orbit& &&&&$i_\text{I}$&&&&&&$39.254(10)^{\circ}$\\
Pulsar mass& &&&&$m_\text{p}$&&&&&&$1.4378(13)  m_{\odot}$\\
Inner companion mass& &&&&$m_\text{I}$&&&&&&$0.19751(15)  m_{\odot}$\\
Outer companion mass& &&&&$m_\text{O}$&&&&&&$0.4101(3)  m_{\odot}$\\

  \noalign{\smallskip}\hline
\end{tabular}
\end{center}
\end{table}

The most important consequence of the Nordvedt effect is a polarization of the Moon's orbit about the Eath. Because the Moon's self-gravitational energy is smaller than the Earth's, the Nordtvedt effect causes the Earth and Moon to fall toward the Sun with slightly different accelerations \cite{Nord68}. Similarly, the triple system, where the pulsar is in a tight orbit with a white-dwarf companion, and this inner binary falls in the gravitational of a third companion. This would resemble the SEP test done in the Earth-Moon-Sun system, but with a strongly self-gravitating object. The polarization of the inner binary system by the external field of the third companion is given in Eq.(5) \cite{Will1985}:
 \begin{align}
\delta r_{\Delta}=\left(\frac{1+2\omega_\text{I}/ \Lambda}{\omega_\text{I}^2-\Lambda^2}\right)\delta a \cos{\Lambda t},
\end{align}
Where $\delta a=\eta \left[\left(E_g/m\right)_P-\left(E_g/m\right)_I\right]Gm_\text{O}/R^2$, $\omega_\text{I}$ is the angular velocity of the pulsar around the inner companion, and $\Lambda=\omega_\text{I}-\omega_\text{O}$ with $\omega_\text{O}$ is the angular velocity around the outer companion.

Submitting the parameters listed in Table I into Eq.(6),  we finally get,
 \begin{align}
\delta r_{\Delta}=1.32851\times 10^{9} \eta \cos{(4.4408\times 10^{-5} t)}\text{m}. \label{drn}
\end{align}
From $r=a(1-e\cos E)$, considering in our case, the inner eccentricity is very small, we can conclude that the uncertain value of the orbital radii is just the one of inner pulsar semi-major axis $\Delta a=4 \times 10^{-4} \text{ls}$ as shown in Table I. \comment{$\Delta e=2\times 10^{-8}$,} Then we get the uncertain orbital radii:
\begin{align}
\Delta r\leq 1.2\times 10^5 \text{m}. \label{dra}
\end{align}
\comment{we can assume that the  $\Delta r=\Delta a(1-e\cos E)-a\Delta e \cos E$ as the variation of the radius of the pulsar orbit around the inner companion, where $\Delta a$ and $\Delta e$ are the uncertain value of inner pulsar semimajor axis $a_I$ and inner eccentricity $e_I$. So that we can estimate the value of $\Delta r$ as:
\begin{align}
\Delta r=\Delta a(1-e\cos E)-a\Delta e \cos E \leq \Delta a+\Delta a e+a\Delta e.
\end{align}
}
The post-Newtonian periastron advance will also produce a variation of the orbital radii. \comment{In this triple system, the advance of the periastron the inner binary , which is a long-term effect.} But a simple calculation shows that this post-Newtonian effect to the radius of the pulsar orbit is only about $7.0$ km, which is much smaller than the uncertain value of $\Delta r$. And also the time scale of the periastron shift is much longer than the outer orbital period. Then we can conclude that $\delta r_{\Delta}$ due to the violation of SEP can not be larger than $\Delta r$ which is from the uncertain value of $a_\text{I}$. We then can constrain the Nordvedt parameter by combining Eq. (\ref{drn}) and Eq. (\ref{dra}):
\begin{align}
 \eta \leq 9.0\times 10^{-5} < 1.0\times 10^{-4}. \label{uplimit1}
\end{align}
It means that from the orbital parameters of PSR J0337+1715, the Nordtvedt parameter must be less than a value of $1\times 10^{-4}$. This estimation is a little smaller than the value given by LLR experiments ($|\eta_\text{N}| = 4.4 \pm 4.5 \times 10^{-4}$) \cite{Will06, Adelberger, Baessler}.

\emph{Discussion and Conclusions}
In this paper we presented new limits on the violation of the SEP in strong field regimes. By analyzing the triple system as the binary pulsars falls in the gravitational field of a third companion, we estimate the Nordvedt parameter $\eta < 10^{-4}$.

The triple system also can be treated another way: the pulsar and its inner white-dwarf companion as the small-eccentricity binary, and the outer white-dwarf companion as the source of gravitational acceleration field $\boldsymbol{g}$. Because of the very small eccentricity, the eccentricity vector $\boldsymbol{e}(t)$ of the pulsar's inner orbit is given by the following vectorial superposition like in the literature \cite{Damour91, Wex12}:
\begin{align}
\boldsymbol{e}(t)=\boldsymbol{e}_{\Delta}(t)+\boldsymbol{e}_\text{R}(t)
\end{align}
And the average variation of eccentricity on time caused by a violation of SEP is
\begin{align}
\dot{e} = <d\boldsymbol{e}/dt> \cdot \hat{\boldsymbol{e}} = \sqrt{1-e^2} (\hat{\boldsymbol{k}} \times \hat{\boldsymbol{e}}) \cdot \boldsymbol{f}. \label{echange}
\end{align}
where $\boldsymbol{e} = e\hat{\boldsymbol{e}}$, $\boldsymbol{f}=\frac{3}{2\mathcal{V}_0}\Delta \boldsymbol{g}$, and $\mathcal{V}_0 = (GM\omega_\text{I})^{1/3}$ is a measure for the relative orbital velocity between the pulsar and its companion; the direction of eccentricity vector is along apsidal line (towards the periastron); $\omega_\text{R}$ is the average angular velocity of the relativistic advance of the periastron.

Since its outer orbital frequency is much larger than the periastron-shift rate of the inner binary, i.e., $\omega_\text{R} = \frac{3(\mathcal{V}_0/c)^2}{1-e^2}\omega_I = 6.77 \times 10^{-11} \text{rad/s} \ll \omega_\text{O}.$ The solution of Eq. (\ref{echange}) is,
\begin{align}
\boldsymbol{e}_\Delta \approx \boldsymbol{f}/2\omega_\text{O}.
\end{align}
In addition, our calculation shows that $|\boldsymbol{e}_{\text{R}}|$ is still less than the uncertain value of $e$ measured about two orders \cite{pulsar3}. So we can assume $|\boldsymbol{e}_\Delta|$ must be not larger than the uncertain value of $e$: $2 \times 10^{-8}$, i.e.,
\begin{align}
|\boldsymbol{e}_\Delta| \approx |\boldsymbol{f}|/2\omega_\text{O} = 0.14 \eta \leq 2 \times 10^{-8},
\end{align}
then we can get a Nordtvedt parameter $\eta \lesssim 10^{-7}$. This result is much better than the one calculated by using the polarization in Eq. (\ref{uplimit1}). But because of the tiny eccentricity itself, this value maybe not very confident.

However, in such a triple system, the challenge lies in obtaining a sufficient number of pulsar time of arrival (TOA) observations to constrain the orbit and the mass of the companion such that  any effects from SEP violation can be separated from any other perturbations. Since the observation time baseline is only 2 years, more timing data of PSR J0337+1715 in future are needed to confirmed this result and to obtain accurate constrain on the Nordtvedt effect. And also, a post-Newtonian dynamical model should replace the Newtonian one used in \cite{pulsar3} for data processing. We are sure that this triple system will play a key role in test of General Relativity.

\emph{Acknowledgments} This work was supported by the NSFC (No.11273045).



\end{document}